\documentclass[aps,prd,superscriptaddress,preprintnumbers,reprint,nofootinbib,showpacs,floatfix]{revtex4-1}

\pdfoutput=1

\usepackage{graphicx}
\usepackage{epsfig}
\usepackage{amssymb}

\usepackage{amsmath}
\usepackage{amsfonts}
\usepackage{fancyhdr}
 
\usepackage{amsmath}
\usepackage{anysize}
\usepackage{verbatim}
\usepackage{cancel} 
\usepackage{framed}
\usepackage[titletoc]{appendix}
\usepackage{mathrsfs}
\usepackage{amsbsy}
\usepackage{color}      
\usepackage{soul}
\usepackage{footmisc}
\usepackage{placeins}

\usepackage{vmargin}

\usepackage{bm}
\usepackage[dvipsnames]{xcolor}

\setpapersize{A4}
\setmargins{2.5cm}       
{1.5cm}                        
{17cm}                      
{23.42cm}                    
{10pt}                           
{1cm}                           
{0pt}                             
{2cm}                           

\def\lQ{\Lambda_{\rm QCD}}

\def\j{{\cal J}}

\def\m{{\cal M}}
\newcommand{\be}{\begin{equation}}
\newcommand{\ee}{\end{equation}}
\newcommand{\bea}{\begin{eqnarray}}
\newcommand{\eea}{\end{eqnarray}}
\newcommand{\nn}{\nonumber}

\begin{document}

\title{Hyperfine splittings of heavy quarkonium hybrids}
\author{Joan Soto} \author{Sandra Tom\`as Valls}
\affiliation{Departament de F\'\i sica Qu\`antica i Astrof\'\i sica and Institut de Ci\`encies del Cosmos, 
Universitat de Barcelona, Mart\'\i $\;$ i Franqu\`es 1, 08028 Barcelona, Catalonia, Spain}

\date{\today}

\preprint{}

\begin{abstract}

In the framework of the Born-Oppenheimer Effective Field Theory, the hyperfine structure of heavy quarkonium hybrids at leading order in the $1/m_Q$ expansion is determined by two potentials. We estimate those potentials by interpolating between the known short distance behavior and the long distance behavior calculated in the QCD Effective String Theory. The long distance behavior depends, at leading order, on two parameters which can be obtained from the long distance behavior of the heavy quarkonium potentials (up to sign ambiguities). The short distance behavior depends, at leading order, on two extra paramentes, which are obtained from a lattice calculation of the lower lying charmonium hybrid multiplets. This allows us to predict the hyperfine splitting both of bottomonium hybrids and of higher multiplets of charmonium hybrids. We carry out a careful error analysis and compare with other approaches.

\end{abstract}

\maketitle

\section{Introduction}\label{sec:intr}

Exotic hadrons (those beyond mesons and baryons) have been a matter of research since the early days of QCD \cite{Jaffe:1975fd}. Among them, the so called hybrids are the closests to standard hadrons, as the only difference resides in their non-trivial gluon content. Hence their flavor structure is the same as in standard hadrons, but their $J^{PC}$ quantum numbers may differ since, in general, the non-trivial gluon content contributes to them. Nevertheless, the experimental confirmation of hybrids is an arduous task. On the one hand, hadrons with exotic quantum numbers, which could be associated to hybrid states, are difficult to produce with current beams. On the other hand, for light hadrons, hybrids with standard quantum numbers may mix with standard hadrons in an arbitrary way. However, when heavy quarks are involved the mixing is suppressed by inverse powers of the heavy quark mass, and hence the identification of hybrids with standard quantum numbers should become simpler, provided we have reliable theoretical predictions for the hybrid spectrum.

An economical approach to calculating the hybrid spectrum is the so-called Born-Oppenheimer effective field theory (BOEFT) \cite{Braaten:2014qka,Berwein:2015vca,Oncala:2017hop,Soto:2017one,Brambilla:2018pyn,Soto:2020xpm}. It exploits the fact that heavy quarks move slowly in heavy hadrons. 
The effect of the non-trivial gluon (or/and light quark) content is encoded in a series of potentials organized in an $1/m_Q$ expansion, $m_Q$ being the heavy quark mass. The leading potential ($\mathcal{O}(1/m_Q^0)$) is heavy quark spin and heavy quark mass independent. It has been used to calculate the spin average spectrum \cite{Juge:1999ie,Braaten:2014qka,Berwein:2015vca,Oncala:2017hop}, decays to heavy quarkonium \cite{Braaten:2014ita,Oncala:2017hop,TarrusCastella:2021pld,Brambilla:2022hhi}, and transitions between heavy quarkonium states \cite{Pineda:2019mhw}. The mixing of heavy quarkonium hybrids with heavy quarkonium starts 
at order $1/m_Q$ \cite{Oncala:2017hop}. Spin effects also start at order $1/m_Q$ \cite{Oncala:2017hop,Soto:2017one}, and hence, they are more important than in heavy quarkonium, in which they start at order $1/m_Q^2$. In that respect, it is important to have spin effects under good control in order to properly identify possible experimental candidates (see \cite{Brambilla:2019esw,Chen:2022asf} for recent reviews). 

We calculate here the hyperfine splittings (HFS) for the lower lying charmonium and bottomonium hybrids at leading-order (LO) ($\mathcal{O}(1/m_Q)$) in the BOEFT. At this order, the HFS depend on two unknown potentials \cite{Soto:2017one,Brambilla:2018pyn}. At short distances the form of these potentials is constrained by the multipole expansion and has been given in ref. \cite{Brambilla:2018pyn,Brambilla:2019jfi}, where one can also find the form of the relevant next-to-leading order (NLO) potentials ($\mathcal{O}(1/m_Q^2)$). The HFS was calculated in that reference using the short distance form of the potentials only. At long distances, the form of the potentials can be estimated using the QCD effective string theory (EST) \cite{Luscher:2002qv,Luscher:2004ib}. This has been carried out for heavy quarkonium \cite{Perez-Nadal:2008wtr,Brambilla:2014eaa} and for the hybrid-quarkonium mixing terms \cite{Oncala:2017hop}. We provide the results here for the spin-dependent terms of the lower lying static hybrid states ($\Sigma_u$ and $\Pi_u$), which turn out to be parameter free (up to signs). We emphasize that the typical distance between heavy quarks in heavy quarkonium hybrids states is of $\mathcal{O}(1/\lQ)$, and hence, an interpolation between the short and long distance forms of the spin-dependent potentials should provide more reliable estimates than sticking to the short distance form only. We propose a simple interpolation and calculate the HFS with it. We use the charmonium spectrum of ref. \cite{Cheung:2016bym} to fix the unknown parameters in the short distance form of the potential, and to estimate the interpolation dependence. Then we can predict the HFS of higher multiplets and the HFS of bottomonium hybrids. 

We distribute the paper as follows. In Section \ref{sec:hpf}, we work out the structure of the spin-dependent terms in a convenient basis.
We specify the form of the spin-dependent potentials at short distances in \ref{sec:sdp}, at long distances in \ref{sec:ldp}, and the interpolation we use in \ref{sec:ip}. In Section \ref{sec:cc}, we fix our free parameters using hybrid charmonium lattice data for the lower multiplets and obtain the HFS of higher multiplets. In Section \ref{sec:bb}, we obtain the HFS of bottomonium hybrids. Section \ref{sec:cp} is devoted to the comparison with other approaches and we close with a discussion and our conclusions in Section \ref{sec:dis}.

\section{The hyperfine splittings}\label{sec:hpf}

General expressions for the BOEFT at NLO have been recently obtained in ref. \cite{Soto:2020xpm}. The lower lying hybrid potentials correspond to the $\kappa^p=1^+$ quantum numbers of the light degrees of freedom (LDOF), where $\kappa$ is the total angular momentum and $p$ the parity. We focus on the heavy quark spin-dependent terms in the Hamiltonian $V_{\kappa^p
}^{(1)}(\mathbf{r})/m_Q$,
\bea
V_{\kappa^p
}^{(1)}(\mathbf{r})&=&\sum_{\Lambda\Lambda'}{\cal P}_{\kappa\Lambda}\left[V^{sa}_{\kappa^p\Lambda\Lambda'}(r)\bm{S}_{Q\bar Q}\cdot\left({\cal P}^{\rm }_{10}\cdot\bm{S}_{\kappa}\right)\right.\nn\\
&+&\left.V^{sb}_{\kappa^p\Lambda\Lambda'}(r)\bm{S}_{Q\bar Q}\cdot\left({\cal P}^{\rm }_{11}\cdot\bm{S}_{\kappa}\right)
\right]{\cal P}_{\kappa\Lambda'}\,,\label{sdp32}
\eea
where $\bm{S}_{Q\bar Q}$ and $\bm{S}_{\kappa}$ are the spin operators of the heavy quark-antiquark pair and the total angular momentum operator of the LDOF respectively. ${\cal P}_{\kappa\Lambda}$ are
the projectors to the irreducible representations of the $D_{h\infty}$ group $\Lambda=0,1$. They read,
\bea
{\cal P}_{10}&=&\mathbb{I}_{3}-\left(\hat{\bm{r}}\cdot\bm{S}_{1}\right)^2\,,\nn\\
{\cal P}_{11}&=&\left(\hat{\bm{r}}\cdot\bm{S}_{1}\right)^2
\eea
In our case, the above potentials act on fields $H^{nm}_1(\mathbf{r},t)$, $n,m=1,2,3$, where the first index corresponds to the total angular momentum of the LDOF and the second to the spin of the heavy quark-antiquark pair. In the Cartesian basis, $\left(\bm{S}_1^i\right)^{jk}=-i\epsilon^{ijk}$, we then have,
\bea
{\cal P}_{10}^{n'n}&=&\hat r^{n'}\hat r^n 
\,,\nn\\
{\cal P}_{11}^{n'n}&=&\delta^{n'n}-\hat r^{n'}\hat r^n 
\eea
Analogously, the heavy quark-antiquark spin operators read $\left(\bm{S}_{Q\bar Q}^i\right)^{m'm}=-i\epsilon_{im'm}$ .
We find that only two independent potentials survive. 
\bea
&&\left[V_{\kappa^p%
}^{(1)}(\bm{r})\right]^{n' m';\,n m}=-V^{sa}_{1^+11}(r)\left(
\delta^{n'm}\delta^{nm'}-\delta^{n'm'}\delta^{nm}
\right) \nn\\
&& +\left(V^{sa}_{1^+11}(r)+V^{sb}_{1^+10}(r)\right)\hat r^i\hat r^j\left(
\delta^{jm}\delta^{n'i}\delta^{nm'}+\right.\nn\\
&&\left.+\delta^{ni}\delta^{jm'}\delta^{n'm} -\delta^{jm'}\delta^{n'i}\delta^{nm}-\delta^{ni}\delta^{jm}\delta^{n'm'}
\right)\nn\\
&& = -2V_{hf}(r) \left(
\delta^{n'm}\delta^{nm'}-\delta^{n'm'}\delta^{nm}
\right) \\
&& -2V_{hf2}(r) \left(\hat r^i\hat r^j-\frac{\delta^{ij}}{3}\right)\left(
\delta^{jm}\delta^{n'i}\delta^{nm'}+\right.\nn\\
&&\left.+\delta^{ni}\delta^{jm'}\delta^{n'm} -\delta^{jm'}\delta^{n'i}\delta^{nm}-\delta^{ni}\delta^{jm}\delta^{n'm'}
\right)\,,\nn
\label{sdpexplicit}
\eea
where we have used that time reversal and Hermiticity imply $V^{sb}_{1^+01}(r)=V^{sb}_{1^+10}(r)\in \mathbb{R}$, and we have defined,
\bea
V_{hf}(r)&=&\frac{1}{6}V^{sa}_{1^+11}(r)-\frac{1}{3}V^{sb}_{1^+10}(r)\nn\\ V_{hf2}(r)&=&-\frac{1}{2}\left(V^{sa}_{1^+11}(r)+V^{sb}_{1^+10}(r)\right)
\label{Vhfdef}
\eea
$V^{sa}_{1^+11}(r)$ and $V^{sb}_{1^+10}(r)$ are defined in \eqref{lattice}\footnote{Note that we take the sign of $V_{1^+11}^{sa}(r)$ opposite to the one in Ref. \cite{Soto:2020xpm}.
\label{sign} }. The structure leading to $V_{hf}$ was first noticed in  \cite{Oncala:2017hop} and the one leading to $V_{hf2}$ was already considered in the short distance analysis of Refs. \cite{Brambilla:2018pyn,TarrusCastella:2019lyq}. We write the field $H^{nm}_1(\mathbf{r},t)$ in the $\vert SLJ\mathcal{J}\mathcal{M}>$  basis, where $S$ and $L$ are the spin and the orbital angular momentum  of the $Q\bar Q$ pair respectively, $J$ the total angular momentum of the LDOF plus the orbital angular momentum of the $Q\bar Q$, and $\mathcal{J}$ and $\mathcal{M}$ the total angular momentum and its third component respectively \cite{Oncala:2017hop} . For a given $\mathcal{J}$, $J=\mathcal{J},\mathcal{J}\pm 1$, $L=J,J\pm 1$ and only $S=1$ is affected by (\ref{sdp32}).
\be
H^{ji}_1({\bf r},t)=\frac{1}{r}\sum_{LJ\j\m} Y^{ji\, LJ}_{\j\m}({\hat {\bf r}}) P^{LJ}_{1\j\m}(r)e^{-iEt} \, ,
\ee 
\bea
Y^{ijLJ}_{\cal J M}({\hat {\bf r}})&=&\sum_{\mu,\nu=0,\pm 1} C(J1{\cal J};{\cal M}-\nu\, \nu)  \\
& \times & C(L1J;\mathcal{M}-\mu-\nu\, \mu) Y^{\mathcal{M}-\mu-\nu}_L ({\hat {\bf r}}) \chi^i_\mu\chi^j_\nu \,,\nn
\eea
where $C(JJ'J'';MM')$ are Clebsch-Gordan coefficients, $Y^{M}_L ({\hat {\bf r}})$ the spherical harmonics  and $\chi^i_\mu$ the spin $1$ eigenvectors. In this basis, (\ref{sdp32}) becomes, for $\mathcal{J}>1$, a $9\times 9$ matrix that splits into a $5\times 5$ box and a $4\times 4$ box. 

The five dimensional box corresponds to the subspace spanned by ($(P_{1\mathcal{JM}}^{--}, P_{1\mathcal{JM}}^{+-}, P_{1\mathcal{JM}}^{00}, P_{1\mathcal{JM}}^{-+}, P_{1\mathcal{JM}}^{++}$), where we use the short hand notation $\pm$ for $J\pm 1$ or $\mathcal{J}\pm 1$ and $0$ for $J$ or $\mathcal{J}$. For the terms proportional to 
 $-2V_{hf}$ it reads \cite{pere,Soto:2017one},
\begin{gather}
\label{Hhf_1}
\begin{pmatrix}
    1 & 0 & 0 & 0 & 0  \\
    0 & -\frac{\mathcal{J}-1}{\mathcal{J}} & \frac{\mathcal{J}+1}{\mathcal{J}}\frac{\sqrt{2\mathcal{J}-1}}{\sqrt{2\mathcal{J}+1}} & 0 & 0  \\
    0 & \frac{\mathcal{J}+1}{\mathcal{J}}\frac{\sqrt{2\mathcal{J}-1}}{\sqrt{2\mathcal{J}+1}} & -\frac{1}{\mathcal{J}(\mathcal{J}+1)} & \frac{\mathcal{J}}{\mathcal{J}+1}\frac{\sqrt{2\mathcal{J}+3}}{\sqrt{2\mathcal{J}+1}} & 0  \\
    0 & 0 & \frac{\mathcal{J}}{\mathcal{J}+1}\frac{\sqrt{2\mathcal{J}+3}}{\sqrt{2\mathcal{J}+1}} & -\frac{\mathcal{J}+2}{\mathcal{J}+1} & 0 \\
    0 & 0 & 0 & 0 & 1 \\
\end{pmatrix}
\end{gather}
and for terms proportional to $-2V_{hf2}$ \cite{pol},
\begin{gather}
\label{Hhf2_1}
-
\begin{pmatrix}
    \frac{2(2-\mathcal{J})}{-3+6\mathcal{J}} & V_2^{--} & V_2^{-0} & 0 & 0 \\
    V_2^{--} & \frac{2(1-\mathcal{J}^2)}{3\mathcal{J}-6\mathcal{J}^2} & V_2^{+-} & 0 & 0  \\
    V_2^{-0} & V_2^{+-} & -\frac{2}{3\mathcal{J}(1+\mathcal{J})} & V_2^{00} & V_2^{+0} \\
    0 & 0 & V_2^{00} & \frac{2\mathcal{J}(2+\mathcal{J})}{9+15\mathcal{J}+6\mathcal{J}^2} & V_2^{-+} \\
    0 & 0 & V_2^{+0} & V_2^{-+} & -\frac{2(3+\mathcal{J})}{9+6\mathcal{J}} \\
\end{pmatrix}
\end{gather}
with,
\begin{gather}
\begin{matrix}
    V_2^{-0}=\frac{\mathcal{J}^2-1}{\sqrt{\mathcal{J}(\mathcal{J}-1)(4\mathcal{J}^2-1)}} \\
    V_2^{--}=\frac{\sqrt{\mathcal{J}-1}}{\sqrt{\mathcal{J}}(2\mathcal{J}-1)}\\
    V_2^{+-}=\frac{\mathcal{J}^2-\mathcal{J}-2}{3\mathcal{J}\sqrt{4\mathcal{J}^2-1}} \\
    V_2^{00}=\frac{\mathcal{J}(\mathcal{J}+3)}{3(\mathcal{J}+1)\sqrt{4\mathcal{J}(\mathcal{J}+2)+3}} \\
    V_2^{-+}=-\frac{\sqrt{2+\mathcal{J}}}{(3+2\mathcal{J})\sqrt{1+\mathcal{J}}} \\
    V_2^{+0}=\mathcal{J}\sqrt{\frac{(2+\mathcal{J})}{(1+\mathcal{J})(1+2\mathcal{J})(3+2\mathcal{J})}}
\nonumber
\end{matrix}
\end{gather}
The four dimensional box corresponds to the subspace spanned by ($(P_{1\mathcal{JM}}^{0-}, P_{1\mathcal{JM}}^{-0}, P_{1\mathcal{JM}}^{+0}, P_{1\mathcal{JM}}^{0+}$). For the terms proportional to 
$-2V_{hf}$ it reads \cite{pere,Soto:2017one},
\begin{gather}
\label{Hhf_2}
\begin{pmatrix}
    \frac{1}{\mathcal{J}} & \frac{\sqrt{\mathcal{J}^2-1}}{\mathcal{J}} & 0 & 0 \\
    \frac{\sqrt{\mathcal{J}^2-1}}{\mathcal{J}} & -\frac{1}{\mathcal{J}} & 0 & 0 \\
    0 & 0 & \frac{1}{\mathcal{J}+1} & \frac{\sqrt{(\mathcal{J}+2)\mathcal{J}}}{\mathcal{J}+1} \\
    0 & 0 & \frac{\sqrt{(\mathcal{J}+2)\mathcal{J}}}{\mathcal{J}+1} & -\frac{1}{\mathcal{J}+1} \\
\end{pmatrix}
\end{gather}
\\
and for terms proportional to $-2V_{hf2}$ \cite{pol},
\begin{gather}
    \label{Hhf2_2}
    -
   \begin{pmatrix}
    \frac{2}{3\mathcal{J}} & V_2^{0--0} & V_2^{0-+0} & 0 \\
    V_2^{0--0} & -\frac{2}{3\mathcal{J}}+\frac{2}{1+2\mathcal{J}} & V_2^{-0+0} & \frac{\mathcal{J}\sqrt{1+\frac{1}{1+\mathcal{J}}}}{1+2\mathcal{J}} \\
    V_2^{0-+0} & V_2^{-0+0} & -\frac{2(2+\mathcal{J})}{3+9\mathcal{J}+6\mathcal{J}^2} & V_2^{+00+} \\
    0 & \frac{\mathcal{J}\sqrt{1+\frac{1}{1+\mathcal{J}}}}{1+2\mathcal{J}} &  V_2^{+00+} & -\frac{2}{3+3\mathcal{J}} \\
    \end{pmatrix} 
\end{gather}
with,
\begin{gather}
\begin{matrix}
    V_2^{0--0}=\frac{\sqrt{\mathcal{J}^2-1}(\mathcal{J}+2)}{3\mathcal{J}(2\mathcal{J}+1)} \\
    V_2^{0-+0}=\sqrt{\frac{\mathcal{J}-1}{\mathcal{J}}}\frac{\mathcal{J}+1}{2\mathcal{J}+1} \\
    V_2^{-0+0}=\frac{\sqrt{\frac{\mathcal{J}}{1+\mathcal{J}}}-\sqrt{\frac{\mathcal{J}+1}{\mathcal{J}}}}{1+2\mathcal{J}} \\
    V_2^{+00+}=\frac{\sqrt{\mathcal{J}(2+\mathcal{J})}(\mathcal{J}-1)}{3(\mathcal{J}+1)(2\mathcal{J}+1)}
\nonumber
\end{matrix}
\end{gather}

If $\mathcal{J}=1$, $P_{1\mathcal{JM}}^{--}$ and $P_{1\mathcal{JM}}^{0-}$ do not exist and the matrices are $7\times 7$. If $J=0$, $P_{1\mathcal{JM}}^{--}, P_{1\mathcal{JM}}^{+-}, P_{1\mathcal{JM}}^{00}, P_{1\mathcal{JM}}^{0-}, P_{1\mathcal{JM}}^{-0}, P_{1\mathcal{JM}}^{+0}$ do not exist and the system is reduced to $3\times 3$ matrices for both potentials.

At first order in perturbation theory, only the diagonal terms and the off-diagonal terms corresponding to $L=J\pm 1,\, L'=J\mp 1$ matter. This is because the $0$-th order wave functions have a single component for $L=J$ and two components for $L=J\pm 1$.
Then the following formulas for the masses $M_{S\mathcal{J}}$ of the spin multiplet $J$, which  are independent of the potentials $V_{hf}$ and $V_{hf2}$, hold,
\be
 \frac{M_{1\, J+1}-M_{0\, J}}{M_{1\, J}-M_{0\, J}}=-J \,,\quad \frac{M_{1\, J-1}-M_{0\, J}}{M_{1\, J}-M_{0\, J}}=J+1\,.
\label{mif}
\ee
They were initially derived in ref. \cite{pere,Soto:2017one} for $V_{hf2}=0$, but it is not difficult to see that they are also fulfilled for $V_{hf2}\not=0$ \cite{pol}.

\section{The potentials}\label{sec:pot}

The potentials $V^{sa}_{1^+11}(r)$ and $V^{sb}_{1^+10}(r)$ can be obtained from suitable operator insertions in the expectation values of Wilson loops \cite{Soto:2020xpm}, and hence, they are amenable to lattice evaluations. However, no such evaluation exists to date. The short distance behavior can be worked out with the help of pNRQCD \cite{Pineda:1997bj,Brambilla:1999xf}. It has been obtained at NLO in the $1/m_Q$ expansion in Refs. \cite{Brambilla:2018pyn,Brambilla:2019jfi}. However, the typical distances at which the static hybrid potentials reach the minimum \cite{Juge:2002br,Bali:2003jq,Capitani:2018rox,Schlosser:2021wnr}, the typical quark-antiquark distance in the bound states \cite{Berwein:2015vca}, and the shape of the wave functions (see plots in Section VI of \cite{Oncala:2017hop}), indicate that most of the time the quark-antiquark separation is of the order of $1/\lQ$. It may then be more appropriated incorporating reliable long distance information rather than calculating higher orders at short distances. We shall do that by using the Effective String Theory of QCD (EST) \cite{Luscher:1980fr,Luscher:2002qv}, which describes well the long distance behavior of the static hybrid potentials \cite{Juge:2002br}, as well as those of the $1/m_Q$ and $1/m_Q^2$ potentials for quarkonium \cite{Perez-Nadal:2008wtr,Oncala:2017hop}, calculated in lattice $SU(3)$ pure Yang-Mills theory. We will then stick to the Cornell model's philosophy of using potentials that interpolate between known short and long distance behavior.

\subsection{The short distance behavior}\label{sec:sdp}

In order to estimate the short distance behavior we use the fact that the $1/m_Q$ potentials are analytic in $\mathbf{r}$ in pNRQCD. This implies that,
\bea
 V_{hf}(r)/m_Q&=& A+\mathcal{O}(r^2)\nn\\ 
V_{hf2}(r)/m_Q&=& B r^2+\mathcal{O}(r^4)\,.
\label{Vhfsd}
\eea
 We shall keep the LO terms only. $A$ and $B$ are unknown real constants, $A=c_F\,k_A/m_Q$ and $B=c_F\,k_B/m_Q$. $k_A \sim \lQ^2$ and $k_B \sim \lQ^4$  can be related to expectation values of operator insertions in Wilson lines \cite{Brambilla:2019jfi}. $c_F=c_F(m_Q)$ is a short distance matching coefficient inherited from NRQCD \cite{Manohar:1997qy,Amoros:1997rx,Grozin:2007fh}. We shall take the next-to-leading logarithmic expression for it,  $c_F(m_c)\equiv c_F(\nu=1\, \text{GeV}, m_c) =1.12155$ and $
c_F(m_b)\equiv c_F(\nu=1\,\text{GeV}, m_b)=0.87897$. The corrections to the leading short distance behavior are suppressed by powers of $\lQ^2 r^2$. We display the potentials (\ref{Vhfsd}) in Fig. \ref{fig:pot}.

\subsection{The long distance behavior}\label{sec:ldp}

The long distance behavior can be estimated using the QCD effective string theory \cite{Luscher:2002qv,Luscher:2004ib}, following the mapping given in ref. \cite{Perez-Nadal:2008wtr}. It has been obtained in \cite{javier}, see Appendix \ref{string}:
\bea
\label{Vhfl}
    \frac{V_{1^+11}^{sa}(r)}{m_Q}&=& -\frac{2 c_F\pi^2 g\Lambda'''}{m_Q\kappa r^3}\equiv V_{ld}^{sa}(r)\nn\\ &&\\
     \frac{V_{1^+10}^{sb}(r)}{m_Q}&=&\mp \frac{2c_F g\Lambda'\pi^2}{m_Q\sqrt{\pi\kappa}}\frac{1}{r^2} \equiv V_{ld}^{sb}(r)\nn
    \eea
The parameters $g\Lambda' \sim \Lambda_{QCD}$ and $g\Lambda''' \sim \Lambda_{QCD}$ also appear in the spin-dependent potentials for heavy quarkonium. They have been obtained in \cite{Oncala:2017hop} from the lattice data of ref. \cite{Koma:2006fw,Koma:2009ws}. Using simple interpolations with the right short and long distance behavior a good fit to data is obtained with the following outcome\footnote{This is so except for the spin-spin potential, which was not used for the extraction of $g\Lambda'$ and $g\Lambda'''$ in Ref. \cite{Oncala:2017hop}}. 
\begin{equation}
\label{lambda}
    g\Lambda' \sim-59\,\text{MeV} \hspace{0.5cm};\hspace{0.5cm} g\Lambda''' \sim \pm 230\,\text{MeV}
    \end{equation}
If only long distance data points and the long distance form of the potentials are used, the values of $\vert g\Lambda'\vert$ and $\vert g\Lambda'''\vert$ are $40\%$ and $35\%$ larger respectively. 
$\kappa\simeq 0.187$ GeV$^{2}$ is the string tension and $c_F$ is the same NRQCD matching coefficient that appears in the short distance behavior. The corrections to the leading long distance behavior of the potentials in (\ref{Vhfl}) are suppressed by factors $1/r^2\lQ^2$. We display the potentials (\ref{Vhfdef}) at long distances in Fig. \ref{fig:pot}.\\ 

\subsection{The interpolating potentials}\label{sec:ip}

\begin{figure*}
\begin{center}
\includegraphics[height=0.25\textheight,width=0.48\textwidth]{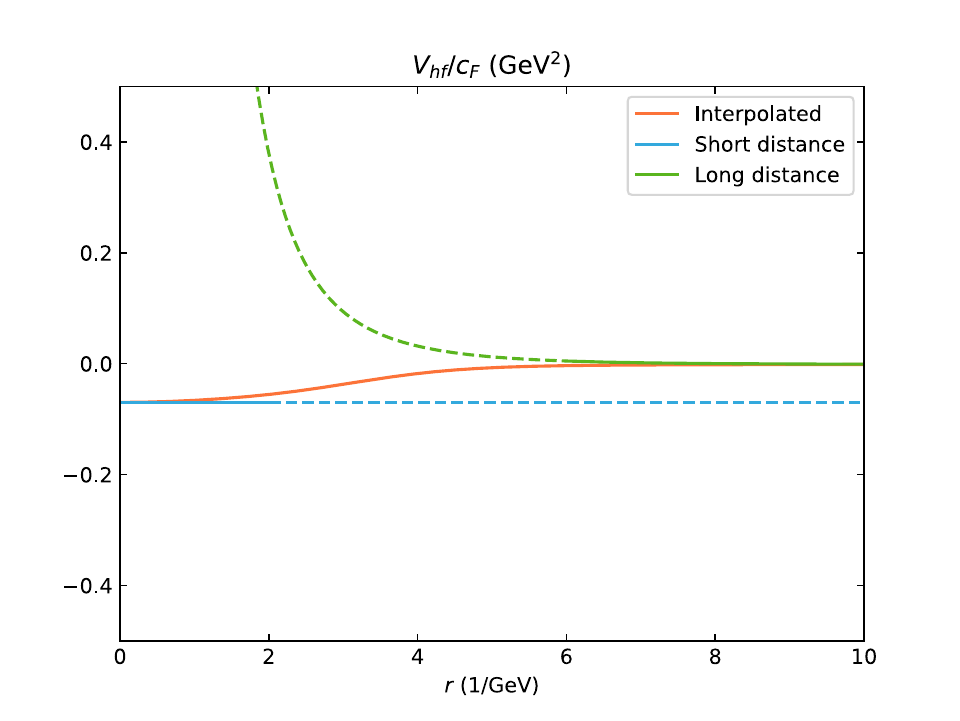}
\includegraphics[height=0.25\textheight,width=0.48\textwidth]{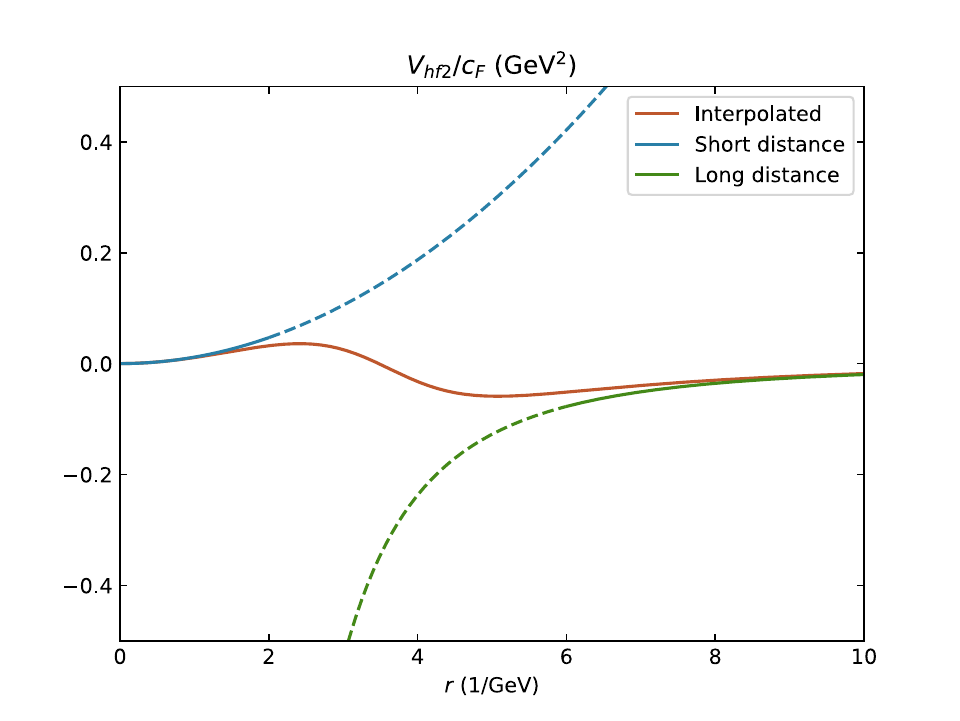} 
\caption{The two spin-dependent potentials (\ref{Vhfdef}) at short (\ref{Vhfsd}) [blue line] and long (\ref{Vhfl}) [green line] distances. The orange line shows the interpolation we use (\ref{Vhf}). The short distance parameters $A$ and $B$ correspond to those in Table \ref{ABerror} and $r_0 = 3.96$ GeV$^{-1}$. }
\label{fig:pot}
\end{center}
\end{figure*}

We use for the hyperfine potentials simple interpolations with the right short and long distance behavior obtained in  Sections \ref{sec:sdp} and \ref{sec:ldp} respectively,
\bea\label{ip}
    \frac{V_{hf}(r)}{m_Q}&=&\frac{A +\left(\frac{r}{r_0}\right)^2\left(\frac{1}{6} V_{ld}^{sa}(r_0)-\frac{r}{3r_0} V_{ld}^{sb}(r_0)\right)}{1+\left(\frac{r}{r_0}\right)^5} \nn\\ &&\\
    \frac{V_{hf2}(r)}{m_Q}&=&\frac{Br^2-\left(\frac{r}{r_0}\right)^5\left(\frac{r_0}{2r} V_{ld}^{sa}(r_0)+\frac{1}{2} V_{ld}^{sb}(r_0)\right)}{1+\left(\frac{r}{r_0}\right)^7}\,. \nn
    \label{Vhf}
\eea 
$r_0 \sim 1/\lQ$ is the matching scale. It is estimated from the short and long distance behavior of the static 
hybrid potentials to be $r_0 \simeq 3.96$ GeV$^{-1}$. This figure will be eventually moved in order to estimate the error due to the interpolation dependence. We show the interpolated potentials in Fig. \ref{fig:pot}.

\section{Charmonium Hybrids: fixing the short distance parameters}\label{sec:cc}

The short distance potentials depend on two arbitrary parameters at LO, $A$ and $B$. We shall fix those parameters by comparing our results to the lattice data of ref. \cite{Cheung:2016bym} for the lower lying hybrid states (the $1(s/d)_1$ ($H_1$), $1p_1$ ($H_2$), $1(p/f)_2$ ($H_4$) and $1p_0$ ($H_3$) multiplets). The spin average of the multiplets in ref. \cite{Cheung:2016bym} were higher than in ref. \cite{Oncala:2017hop}. Since we are going to use the same methodology as in the last reference, we correct the lattice data by the spin average difference, namely by $381$ MeV ($1(s/d)_1$ ), $326$ MeV ($1p_1$), $392$ MeV ($1(p/f)_2$) and $151$ ($1p_0$) MeV. We then scan natural values of $A$ and $B$ in $[-0.3\,, 0.3]$ GeV and $[-0.06\,, 0.06]$ GeV$^3$ respectively and search for the ones with the lowest $\chi^2/dof$. We adapted the code used in ref. \cite{Oncala:2017hop} by adding the spin dependent potentials above, and neglecting, for simplicity, the mixing with quarkonium \cite{sandratfg}. The charm mass is also taken as in ref. \cite{Oncala:2017hop}, $m_c=1.47$ GeV.  If we neglect the long distance behavior, we find $A=-0.0426$ GeV, $B=-0.0014$ GeV$^3$ with $\chi^2/dof=0.806$. When we include the long distance behavior the fit quality improves considerably, we obtain as a best fit  $A=-0.0700$ GeV, $B=0.0117$ GeV$^3$ with $\chi^2/dof=0.582
$. $g\Lambda'$ and $g\Lambda'''$ are taken as in (\ref{lambda}), $r_0 = 3.96$ GeV$^{-1}$ in (\ref{ip}) and all possible sign combinations for the long-distance potentials in (\ref{Vhfl}) and (\ref{lambda}) are considered. The best fit corresponds to a positive $V_{ld}^{sa}(r)$ and $V_{ld}^{sb}(r)$. The former implies $g\Lambda'''<0$, and the later selects the negative sign in the second formula of \eqref{Vhfl}. Reversing the sign of $V_{ld}^{sa}(r)$ ($V_{ld}^{sb}(r)$) worsens the fit considerably (marginally), see Table \ref{tab:signs}.  
 
\begin{table}
    \centering
    \begin{tabular}{|c|c|c|c|c|c|}
    \hline
\multicolumn{2}{|c}{sign($V_{ld}^{sa}$) sign($V_{ld}^{sb}$)} \vline & $+-$ & $++$ & $-+$ & $--$ \\ \hline
\multicolumn{2}{|c}{$\chi^2/dof$} \vline & 0.6471 & 0.5818 & 0.8968 & 0.7418 \\ \hline
\multicolumn{2}{|c}{$A$ (GeV)} \vline & -0.0858 & -0.0700 & -0.0257 & -0.0448 \\ \hline
\multicolumn{2}{|c}{$B$ (GeV$^3$)} \vline & 0.0098 & 0.0117 & -0.0100 & -0.0123 \\ \hline28
$(s/d)_1$ & $1^{--}$ & 4.0107 & 4.0107 & 4.0107 & 4.0107 \\ 
mass & $0^{-+}$ & 3.8862 & 3.8917 & 3.9041 & 3.8985 \\ 
(GeV) & $1^{-+}$ & 3.9513 & 3.9526 & 3.9582 & 3.9556 \\ 
~ & $2^{-+}$ & 4.0593 & 4.0617 & 4.0602 & 4.0621 \\ \hline
$p_1$ mass & $1^{++}$ & 4.1450 & 4.1450 & 4.1450 & 4.1450 \\ 
(GeV) & $0^{+-}$ & 4.1018 & 4.1012 & 4.1046 & 4.1051 \\ 
~ & $1^{+-}$ & 4.1140 & 4.1089 & 4.0920 & 4.1014 \\ 
~ & $2^{+-}$ & 4.1510 & 4.1478 & 4.1396 & 4.1439 \\ \hline
$(p/f)_2$ & $2^{++}$ & 4.2316 & 4.2316 & 4.2316 & 4.2316 \\ 
mass & $1^{+-}$ & 4.2032 & 4.1906 & 4.1662 & 4.1767 \\ 
(GeV) & $2^{+-}$ & 4.2260 & 4.2248 & 4.2301 & 4.2301 \\ 
~ & $3^{+-}$ & 4.2466 & 4.2584 & 4.2791 & 4.2695 \\ \hline
$p_0$ mass & $0^{++}$ & 4.4864 & 4.4864 & 4.4864 & 4.4864 \\ 
(GeV) & $1^{+-}$ & 4.4627 & 4.4773 & 4.4592 & 4.4461 \\ \hline
    \end{tabular}
\caption{Fit parameters and spectrum dependence on the sign ambiguities arising from the long distance spin dependent potentials. }
\label{tab:signs}
\end{table}

We have also explored the dependence of the result on $g\Lambda'$ and $g\Lambda'''$ according to the possible values given in ref. \cite{Oncala:2017hop}.  The fit has a mild preference for smaller absolute values of $\vert g\Lambda'\vert$ and $\vert g\Lambda'''\vert$, so we continue working with $ g\Lambda'=-0.059$ GeV and $ g\Lambda'''=-0.230$ GeV. 
The interpolation dependence is estimated by moving $r_0 \in [2.5\,, 5]$ GeV$^{-1}$. The best $\chi^2/dof$ corresponds to the default value $r_0= 3.96$ GeV$^{-1}$ and considerably deteriorates for $r_0 \le 3$ GeV$^{-1}$, see Table \ref{r0dep}.

\begin{table}
    \centering
    \begin{tabular}{|c|c|c|c|c|c|c|}
    \hline
\multicolumn{2}{|c}{$r_0$ (GeV$^{-1}$)} \vline & $2.5$ & $3$ & $3.5$ & $3.964$ & $5$\\ \hline
\multicolumn{2}{|c}{$\chi^2/dof$} \vline & 1.4686 & 0.7943 & 0.6148 & 0.5818 & 0.6266 \\ \hline
\multicolumn{2}{|c}{$A$ (GeV)} \vline & -0.1691 & -0.1133 & -0.0846 & -0.0700 & -0.0546 \\ \hline
\multicolumn{2}{|c}{$B$ (GeV$^3$)} \vline & 0.1723 & 0.0618 & 0.0253 & 0.0117 & 0.0020 \\ \hline
$(s/d)_1$ & $1^{--}$ & 4.0107 & 4.0107 & 4.0107 & 4.0107 & 4.0107 \\ 
mass & $0^{-+}$ & 3.8756 & 3.8821 & 3.8879 & 3.8917 & 3.8973 \\ 
(GeV) & $1^{-+}$ & 3.9470 & 3.9490 & 3.9511 & 3.9526 & 3.9550 \\ 
~ & $2^{-+}$ & 4.0329 & 4.0513 & 4.0588 & 4.0617 & 4.0628 \\ \hline
$p_1$ mass & $1^{++}$ & 4.1450 & 4.1450 & 4.1450 & 4.1450 & 4.1450 \\ 
(GeV) & $0^{+-}$ & 4.1036 & 4.1008 & 4.1001 & 4.1012 & 4.1040 \\ 
~ & $1^{+-}$ & 4.1255 & 4.1189 & 4.1129 & 4.1089 & 4.1032 \\ 
~ & $2^{+-}$ & 4.1362 & 4.1468 & 4.1486 & 4.1478 & 4.1450 \\ \hline
$(p/f)_2$ & $2^{++}$ & 4.2316 & 4.2316 & 4.2316 & 4.2316 & 4.2316 \\ 
mass & $1^{+-}$ & 4.2253 & 4.2092 & 4.1983 & 4.1906 & 4.1798 \\ 
(GeV) & $2^{+-}$ & 4.2205 & 4.2189 & 4.2221 & 4.2248 & 4.2273 \\ 
~ & $3^{+-}$ & 4.2241 & 4.2403 & 4.2512 & 4.2584 & 4.2681 \\ \hline
$p_0$ mass & $0^{++}$ & 4.4864 & 4.4864 & 4.4864 & 4.4864 & 4.4864 \\ 
(GeV) & $1^{+-}$ & 4.4638 & 4.4865 & 4.4816 & 4.4773 & 4.4754 \\ \hline
    \end{tabular}
\caption{Fit parameters and spectrum dependence on $r_0$. }
\label{r0dep}
\end{table}

In order to establish the error of $A$ and $B$ due to the input data,  we assume that the lattice data errors are fully correlated and performed fits to average, maximum, and minimum values. We obtain $A=-0.070\pm 0.010$ GeV and $B=0.0117\pm 0.0003$ GeV$^3$. The spectrum is displayed in Table \ref{ABerror}. The error due to the uncertainty in $g\Lambda'$ and $g\Lambda'''$ is negligible. We also neglect the error due to the interpolation, which is not negligible in order to obtain a value for $A$ and $B$, but it is for the spectrum due to its correlation with $r0$. We include the error due to higher orders in the $1/m_Q$ expansion, which is about $30$ MeV for charm.

\begin{table}
    \centering
    \begin{tabular}{|c|c|c|c|c|}
    \cline{4-5}
\multicolumn{3}{c}{} \vline & Fit error & Total error \\ \hline
\multicolumn{2}{|c}{$A$ (GeV)} \vline & -0.070 & 0.010 & \\ \hline
\multicolumn{2}{|c}{$B$ (GeV$^3$)} \vline & 0.0117 & 0.0003 & \\ \hline
$(s/d)_1$ & $1^{--}$ & 4.011 &  & 0.030 \\ 
mass & $0^{-+}$ & 3.892 & 0.021 & 0.036 \\ 
(GeV) & $1^{-+}$ & 3.953 & 0.010 & 0.032 \\ 
~ & $2^{-+}$ & 4.062 & 0.009 & 0.031 \\ \hline
$p_1$ mass & $1^{++}$ & 4.145 &  & 0.030 \\ 
(GeV) & $0^{+-}$ & 4.101 & 0.009 & 0.031 \\ 
~ & $1^{+-}$ & 4.109 & 0.010 & 0.032 \\ 
~ & $2^{+-}$ & 4.148 & 0.002 & 0.030 \\ \hline
$(p/f)_2$ & $2^{++}$ & 4.232 &  & 0.030 \\ 
mass & $1^{+-}$ & 4.191 & 0.007 & 0.031 \\ 
(GeV) & $2^{+-}$ & 4.225 & 0.002 & 0.030 \\ 
~ & $3^{+-}$ & 4.258 & 0.005 & 0.030 \\ \hline
$p_0$ mass & $0^{++}$ & 4.486 &  & 0.030 \\ 
(GeV) & $1^{+-}$ & 4.477 & 0.004 & 0.030 \\ \hline
    \end{tabular}
\caption{Fit errors in $A$, $B$ and in the hybrid charmonium spectrum. The total errors in the spectrum are obtained by adding in quadrature the error due to missing higher orders in the $1/m_Q$ expansion ($\sim \lQ^3/m_Q^2\sim 30$ MeV).}
\label{ABerror}
\end{table}

In ref. \cite{Oncala:2017hop}, it was found that two extra multiplets lie below the $1p_0$, the $2(s/d)_2)$ and the $d_2$. 
For completeness we also display the spectrum of these multiplets including the hyperfine splitting in Table \ref{specc}.

\begin{table}
    \centering
    \begin{tabular}{|c|c|c|c|c|}
    \cline{3-5}
\multicolumn{2}{c}{} \vline & Mass (GeV) & $A$ and $B$ error & Total error \\ \hline
$d_2$ & $2^{--}$ & 4.334 &  & 0.030 \\ 
      & $1^{-+}$ & 4.325 & 0.002 & 0.030 \\ 
      & $2^{-+}$ & 4.325 & 0.002 & 0.030 \\ 
      & $3^{-+}$ & 4.334 & 0.0001 & 0.030 \\ \hline
$2(s/d)_1$ & $1^{--}$ & 4.355 &  & 0.030 \\ 
           & $0^{-+}$ & 4.316 & 0.010 & 0.032 \\ 
           & $1^{-+}$ & 4.338 & 0.006 & 0.031 \\ 
           & $2^{-+}$ & 4.374 & 0.004 & 0.030 \\ \hline        
\end{tabular}	
\caption{The remaining hybrid charmonium spectrum below the $1p_0$ multiplet. The total errors in the spectrum are obtained by adding in quadrature to the errors induced by the uncertainties in $A$ and $B$ the error due to missing higher orders in the $1/m_Q$ expansion ($\sim \lQ^3/m_Q^2\sim 30$ MeV). }
\label{specc}
\end{table}

\section{Bottomonium Hybrids: predicting the hyperfine splittings}\label{sec:bb}

Once the parameters $A$ and $B$ are fixed from charmonium, the corresponding parameters for bottomonium, $A'$ and $B'$ are,
\be
A'=A\frac{c_F(m_b) m_c}{c_F(m_c) m_b}\quad B'=B\frac{c_F(m_b) m_c}{c_F(m_c) m_b} \, .
\label{cb}
\ee
We calculate the spectrum for the central values of these parameters $A'=-0.017\pm0.002$ GeV and $B'=0.0028\pm0.0001$ GeV$^3$, which provides us with the central value of the masses, and for the four corners of their $1\,\sigma$ range, which allows us to estimate the error due to the fit parameters. We take it as the larger difference in either sense. 
The total error is obtained by adding in quadrature to the latter the error associated to higher orders $\sim \lQ^3/m_Q^2\sim 3$ MeV. The bottom mass is taken as in ref. \cite{Oncala:2017hop}, $m_b=4.88$ GeV. We display the results in Table \ref{specb}.
\begin{table}
    \centering
    \begin{tabular}{|c|c|c|c|c|}
    \cline{3-5}
\multicolumn{2}{c}{} \vline & Mass (GeV) & $A$ and $B$ error & Total error \\ \hline
$(s/d)_1$ & $1^{--}$ & 10.690 &  & 0.003 \\ 
         & $0^{-+}$ & 10.680 & 0.0021 & 0.004 \\ 
         & $1^{-+}$ & 10.685 & 0.0010 & 0.003 \\ 
         & $2^{-+}$ & 10.695 & 0.0007 & 0.003 \\ \hline
$p_1$ & $1^{++}$ & 10.761 &  & 0.003 \\ 
         & $0^{+-}$ & 10.755 & 0.0011 & 0.003 \\ 
         & $1^{+-}$ & 10.758 & 0.0007 & 0.003 \\ 
         & $2^{+-}$ & 10.764 & 0.0002 & 0.003 \\ \hline
$(p/f)_2$ & $2^{++}$ & 10.819 &  & 0.003 \\ 
         & $1^{+-}$ & 10.814 & 0.0011 & 0.003 \\ 
         & $2^{+-}$ & 10.817 & 0.0003 & 0.003 \\ 
         & $3^{+-}$ & 10.822 & 0.0005 & 0.003 \\ \hline
$d_2$ & $2^{--}$ & 10.870 &  & 0.003 \\ 
         & $1^{-+}$ & 10.868 & 0.0005 & 0.003 \\ 
         & $2^{-+}$ & 10.869 & 0.0003 & 0.003 \\ 
         & $3^{-+}$ & 10.871 & 0.0001 & 0.003 \\ \hline
$2(s/d)_1$ & $1^{--}$ & 10.885 &  & 0.003 \\ 
         & $0^{-+}$ & 10.880 & 0.0012 & 0.003 \\ 
         & $1^{-+}$ & 10.883 & 0.0006 & 0.003 \\ 
         & $2^{-+}$ & 10.888 & 0.0004 & 0.003 \\ \hline
$2p_1$ & $1^{++}$ & 10.970 &  & 0.003 \\ 
         & $0^{+-}$ & 10.967 & 0.0008 & 0.003 \\ 
         & $1^{+-}$ & 10.968 & 0.0005 & 0.003 \\ 
         & $2^{+-}$ & 10.971 & 0.0002 & 0.003 \\ \hline
$2(p/f)_2$ & $2^{++}$ & 11.005 &  & 0.003 \\ 
         & $1^{+-}$ & 11.002 & 0.0007 & 0.003 \\ 
         & $2^{+-}$ & 11.004 & 0.0002 & 0.003 \\ 
         & $3^{+-}$ & 11.007 & 0.0003 & 0.003 \\ \hline
$p_0$ & $0^{++}$ & 11.012 &  & 0.003 \\ 
         & $1^{+-}$ & 11.012 & 0.0001 & 0.003 \\ \hline
    \end{tabular}		
\caption{The hybrid bottomonium spectrum. The total errors in the spectrum are obtained by adding in quadrature to the fit errors the error due to missing higher orders in the $1/m_Q$ expansion ($\sim \lQ^3/m_Q^2\sim 3$ MeV). }
\label{specb}
\end{table}

\section{Comparison with other approaches}\label{sec:cp}

We shall compare here our results with those also obtained in BOEFT in Refs. \cite{Brambilla:2018pyn,Brambilla:2019jfi} and with the lattice QCD results of the HSC collaboration \cite{Cheung:2016bym,Ryan:2020iog} (we will not use earlier results at larger pion masses \cite{HadronSpectrum:2012gic}). Recall that the main differences with respect to Refs. \cite{Brambilla:2018pyn,Brambilla:2019jfi} is that only short distance expressions for the spin-dependent potentials at NLO ($8$ non-perturbative parameters to fit) are used there, whereas we use LO short distance expressions and LO long distance expressions ($2$ non-perturbative parameters to fit). 

\begin{figure*}
\begin{center}
\includegraphics[height=0.25\textheight,width=0.48\textwidth]{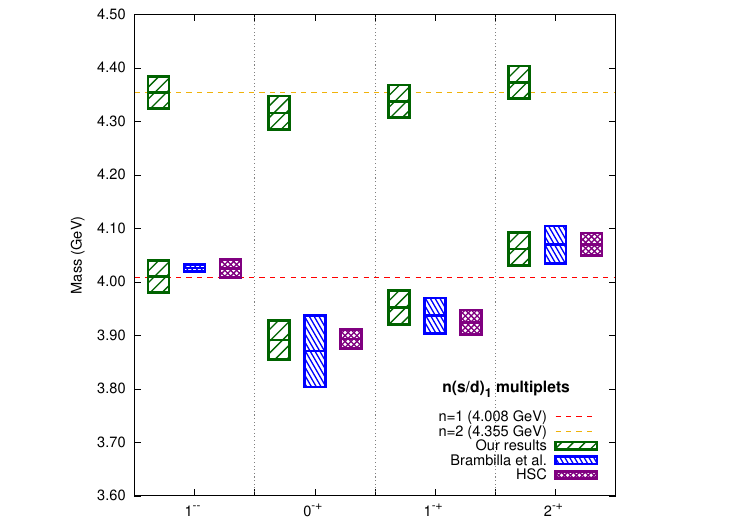}
\includegraphics[height=0.25\textheight,width=0.48\textwidth]{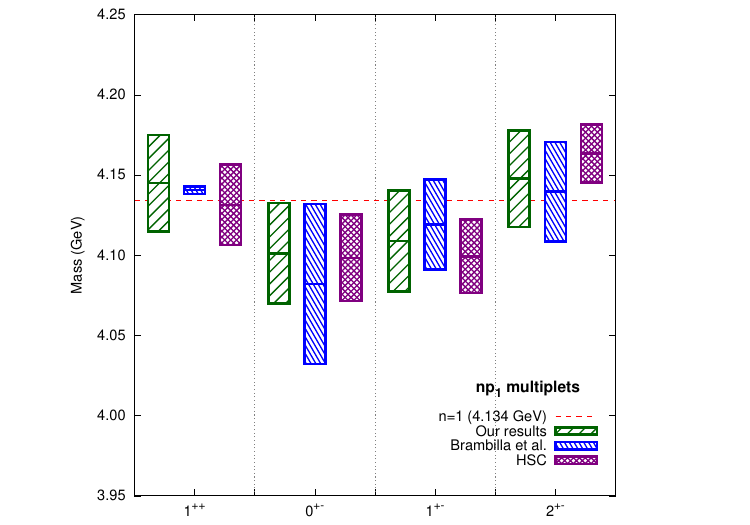} 
\\[2ex]
\includegraphics[height=0.25\textheight,width=0.48\textwidth]{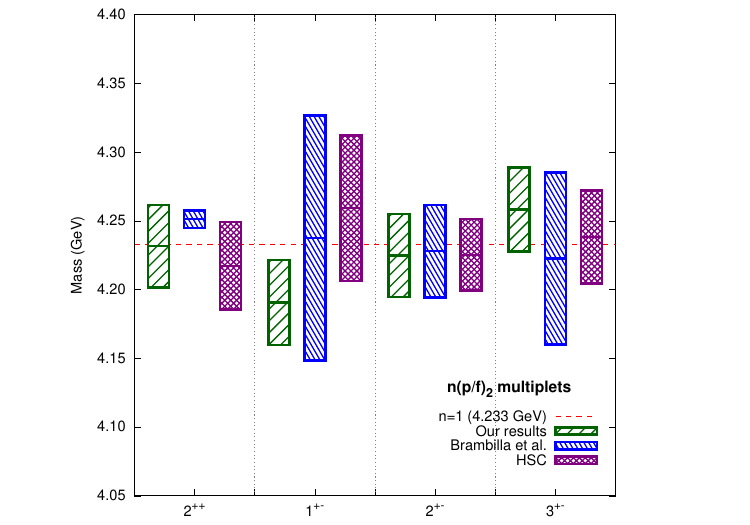}
\includegraphics[height=0.25\textheight,width=0.48\textwidth]{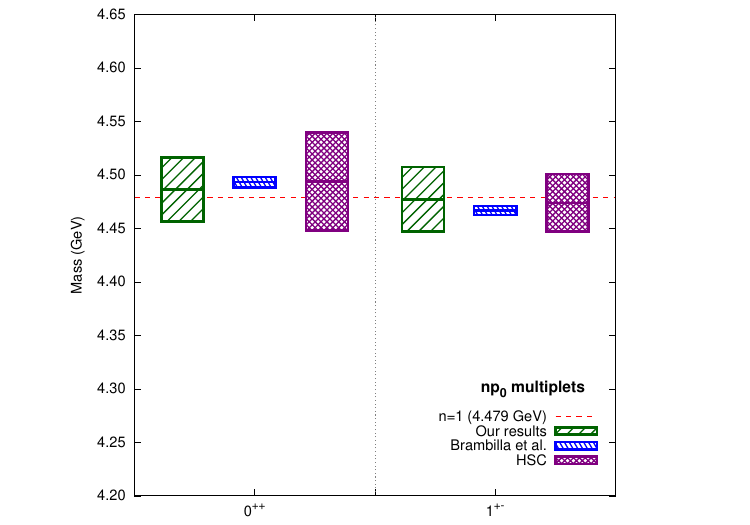}
\caption{The spectrum of the lower-lying $n\,(s/d)_1$ ($H_1$), $n\,p_1$ ($H_2$), $n\,(p/f)_2$ ($H_4$) and $n\,p_0$ ($H_3$) charmonium hybrids computed by adding the LO spin-dependent potentials 
to the static potentials used in ref.~\cite{Oncala:2017hop} is shown in green boxes. 
The average mass for each multiplet is shown as a red dashed line. Blue and magenta boxes show the results of the BOEFT with NLO spin-dependent short distance potentials only \cite{Brambilla:2018pyn,Brambilla:2019jfi} and lattice QCD \cite{Cheung:2016bym} respectively, adjusted to our spin average. 
 The height of the boxes indicates the uncertainty.}
\label{fig:ccg}
\end{center}
\end{figure*}

Lattice results for charmonium hybrids \cite{Cheung:2016bym} are used to fit the non-perturbative parameters. Hence, our errors are slightly larger than the ones from lattice data. Our fits have lower $\chi^2/{\rm dof}$ than those in Refs. \cite{Brambilla:2018pyn,Brambilla:2019jfi}, which supports the inclusion of long distance potentials. The overall picture for the charmonium hyperfine splittings is similar to the one obtained in those references, see Fig. \ref{fig:ccg}. For spin one states we get similar, slightly smaller, smaller and larger errors for the $(s/d)_1$ ($H_1$), $p_1$ ($H_2$), $(p/f)_2$ ($H_4$) and $p_0$ ($H_3$) multiplets respectively. For spin zero states, which do not enter in our analysis, the errors are only due to neglected higher orders in the $1/m_Q$ expansion. Note that the error that we assign to them is larger than than the one assigned in Refs. \cite{Brambilla:2018pyn,Brambilla:2019jfi}. See Fig. \ref{fig:ccg}. 

\begin{figure*}
\begin{center}
\includegraphics[height=0.25\textheight,width=0.480\textwidth]{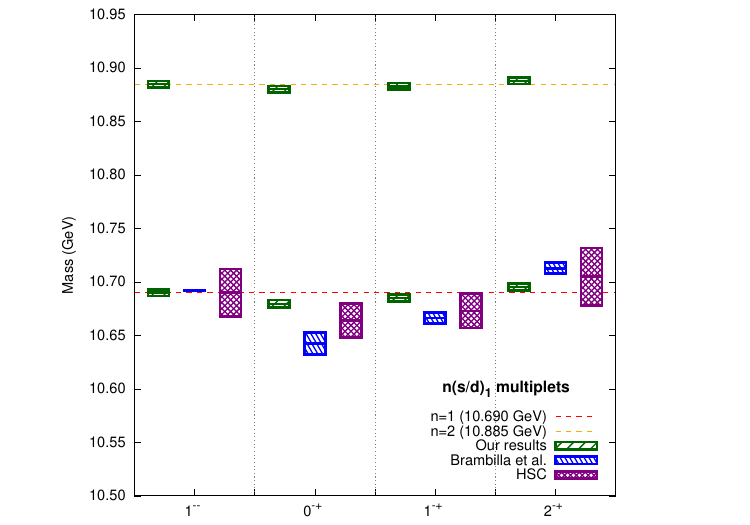}
\includegraphics[height=0.25\textheight,width=0.480\textwidth]{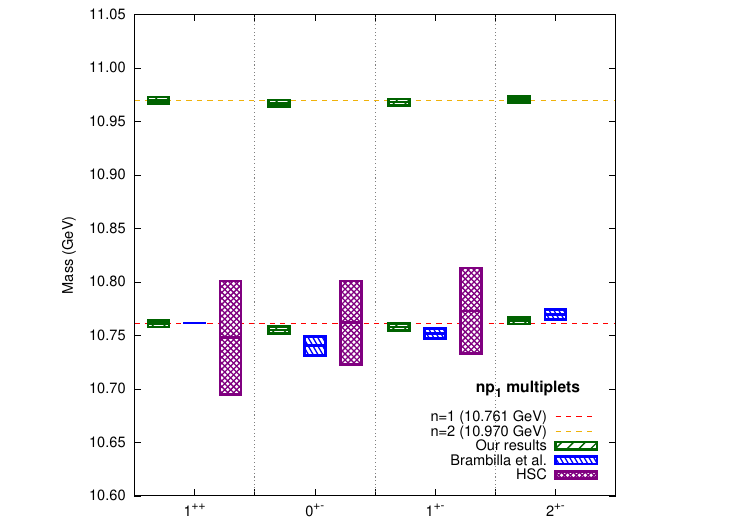} 
\\[2ex]
\includegraphics[height=0.25\textheight,width=0.480\textwidth]{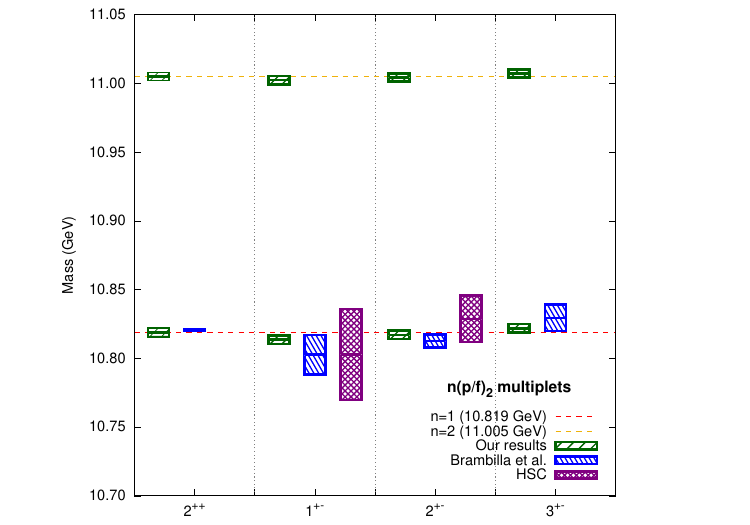}
\includegraphics[height=0.25\textheight,width=0.480\textwidth]{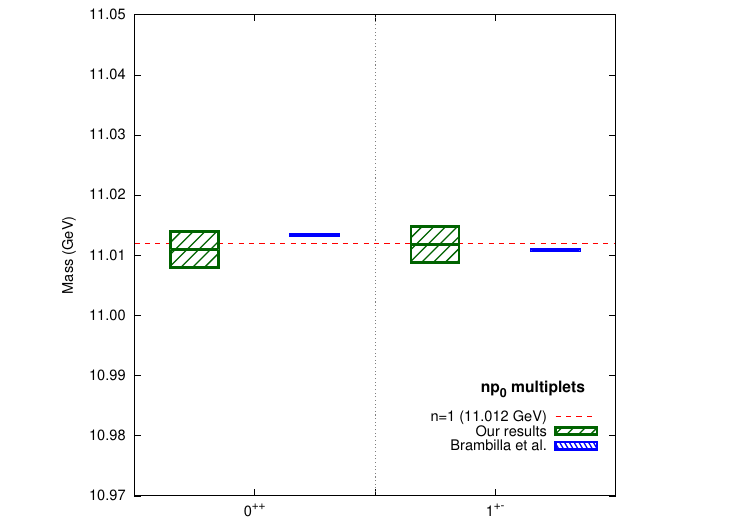}
\caption{The spectrum of the lower-lying $n\,(s/d)_1$ ($H_1$), $n\,p_1$ ($H_2$), $n\,(p/f)_2$ ($H_4$) and $n\,p_0$ ($H_3$) bottommonium hybrids computed by adding the LO spin-dependent potentials 
to the static potentials used in ref.~\cite{Oncala:2017hop} is shown in green boxes. 
The average mass for each multiplet is shown as a red dashed line. Blue and magenta boxes show the results of the BOEFT with NLO spin-dependent short distance potentials only \cite{Brambilla:2018pyn,Brambilla:2019jfi} and lattice QCD \cite{Ryan:2020iog} respectively, adjusted to our spin average. 
 The height of the boxes indicates the uncertainty.}
\label{fig:bbg}
\end{center}
\end{figure*}

Our predictions for the bottomonium hybrid hyperfine splittings are compatible with the few available states from lattice QCD \cite{Ryan:2020iog}, but we have much smaller errors, see Fig. \ref{fig:bbg}. We get again an overall picture similar to the one of ref. \cite{Brambilla:2018pyn,Brambilla:2019jfi}, with smaller errors for spin one states, and smaller hyperfine splittings in general. However, for the $(s/d)_1$ multiplet we have about $2\sigma$ discrepancies with those references.
Nevertheless, both sets of splittings are consistent with lattice data of \cite{Ryan:2020iog} for this multiplet. The remaining available lattice states do not form any obvious spin multiplet. We have assigned the three lighter (two heavier) ones to the $p_1$ ($(p/f)_2$) multiplet. However, the fact that the $1^{++}$ state is lighter than the $1^{--}$ in the lattice results is in conflict with the BOEFT ones. 


\section{Discussion and Conclusions}\label{sec:dis}

The static potentials we use are taken from \cite{Oncala:2017hop}. They were obtained from pure $SU(3)$ ($n_f=0$) lattice data of Refs. \cite{Juge:2002br,Bali:2003jq}. More recent pure $SU(3)$ ($n_f=0$) lattice data exist for the static hybrid potentials, with better resolution at short and intermediate distances \cite{Capitani:2018rox,Muller:2019joq,Schlosser:2021wnr}. It would be interesting to incorporate it in future analysis. However, the systematic errors due to using pure $SU(3)$ ($n_f=0$) rather than QCD with dynamical light quarks ($n_f=3$) would still be difficult to evaluate. The early study of ref. \cite{Bali:2000vr} suggest that they are small. The strategy recently presented in \cite{DallaBrida:2022eua} may help to resolve this issue.

The lattice data to which we fit our spin-dependent expressions, ref. \cite{Cheung:2016bym}, is obtained from a $2+1+1$ clover action at a fixed spacial lattice spacing of $0.12$ fm, a $3.5$ times smaller temporal lattice spacing, and light quark masses corresponding to $m_\pi\sim 240$ MeV. In ref. \cite{Ray:2021nhe} results in the continuum limit for realistic light quark masses are obtained using a HISQ action with $2+1+1$ dynamical quarks, but only for two states in the lowest lying multiplet. The lattice bottomonium data we compare with ref. \cite{Ryan:2020iog} uses the same setting as \cite{Cheung:2016bym}, tuning the heavy quark parameters to bottomonium observables, and uses light quark masses corresponding to $m_\pi\sim 391$ MeV.

We have focused on the hyperfine splittings. The absolute values of the masses we quote correspond to the spin averages of Ref. \cite{Oncala:2017hop}. The central values quoted in that reference are lower than those in Ref. \cite{Berwein:2015vca} but compatible within errors. The differences are due to the choice of normalization (quarkonium spectrum versus RS mass scheme). They are also much lower than the lattice results of Ref. \cite{Cheung:2016bym}, a difference that shrinks if they are compared with earlier lattice data at larger pion mass \cite{HadronSpectrum:2012gic}. 
This suggest that part of the discrepancy may be due to the quenched lattice data used as an input in Refs. \cite{Berwein:2015vca,Oncala:2017hop}. They are also lower than in most models (see \cite{Meyer:2015eta} for a review). Since the discrepancies usually amount to global shifts, they are not expected to affect the bulk of the hyperfine splitting analysis presented here. However, a small dependence on the input lattice data (unphysical) pion mass was noticed in the short distance analysis of Ref. \cite{Brambilla:2019jfi}, which may be present in our results as well.

We have shown that the inclusion of long distance contributions calculated in the QCD EST to the LO spin dependent potentials in the BOEFT considerable improves the description of the charmonium hybrids hyperfine splittings obtained in lattice QCD \cite{Cheung:2016bym}. For the LO spin-dependent potentials, the $\chi^2/dof$ moves from  $0.806$ for short distance contributions only to $0.582$ for short and long distance contributions. This figure is much lower than the $\chi^2/dof=0.999$ obtained in ref. \cite{Brambilla:2019jfi} for NLO spin-dependent potential with short distance contributions only. The fact that long distance contributions are important may be anticipated from the results on the size of charmonium hybrids displayed in Table III of ref. \cite{Berwein:2015vca}, $\langle  1/r\rangle \in \left[190,420\right]$ MeV, scales of the order of $\lQ$. Using the QCD EST to describe them has the remarkable feature that it introduces no new unknown parameter, beyond sign ambiguities and the scale $r_0$ that separates short and long distances in the interpolation. Hence, we have two parameter fits to data, rather than the eight parameter fits of ref. \cite{Brambilla:2019jfi}, leading to a smaller $\chi^2/dof$ as mentioned above.

Once we have the unknown parameters fixed, we can calculate the hyperfine splittings of higher charmonium hybrid states, of the bottomonium ones, and the error associated to them. This is displayed in Tables \ref{ABerror}, \ref{specc} and \ref{specb} and in Figs. \ref{fig:ccg} and \ref{fig:bbg}. For charmonium hybrids, we get results compatible with ref. \cite{Brambilla:2019jfi} with similar errors overall, and, as expected, compatible with ref. \cite{Cheung:2016bym}, the source of our fit, with slightly larger errors. For bottomonium hybrids, our hyperfine splittings are compatible with those of Ref. \cite{Ryan:2020iog}, with much smaller errors, but smaller than those in Ref. \cite{Brambilla:2019jfi}, with similar errors.

\begin{acknowledgments}

We thank Rub\'en Oncala for providing us with the code used in Ref. \cite{Oncala:2017hop}, and Jaume Tarr\'us Castell\`a and Jorge Segovia for providing us for the data and figures used in Refs. \cite{Brambilla:2018pyn,Brambilla:2019jfi,TarrusCastella:2019lyq}. 
 J.S. acknowledges financial support from Grant No. 2017-SGR-929 and 2021-SGR-249 from the Generalitat de Catalunya and from projects No. PID2019- 105614GB-C21, No. PID2019-110165GB-I00 and No. CEX2019-000918-M from Ministerio de Ciencia, Innovaci\'on y Universidades. S.T.V. acknowledges financial support from the department's collaboration grant call 2022-2023 bestowed by Ministeri d'Eduaci\'o i Formaci\'o Professional.

 \end{acknowledgments}

\vspace{0.5cm}
\centerline  {\bf  ADDENDUM}
\vspace{0.5cm}

\begin{figure*}[!t]
\begin{center}
\includegraphics[height=0.25\textheight,width=0.48\textwidth]{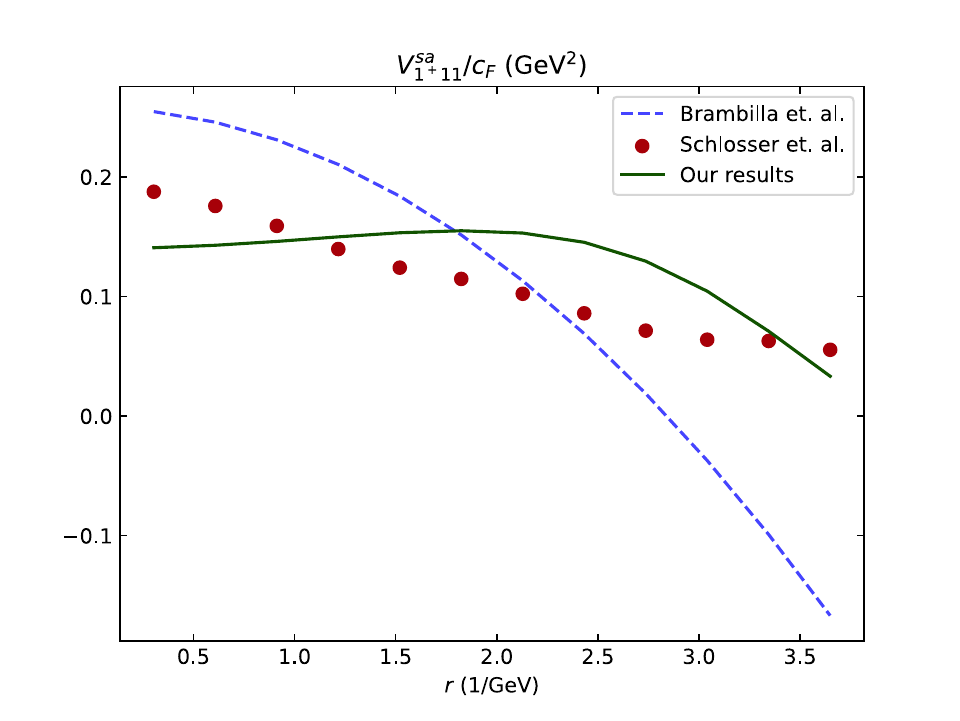}
\includegraphics[height=0.25\textheight,width=0.48\textwidth]{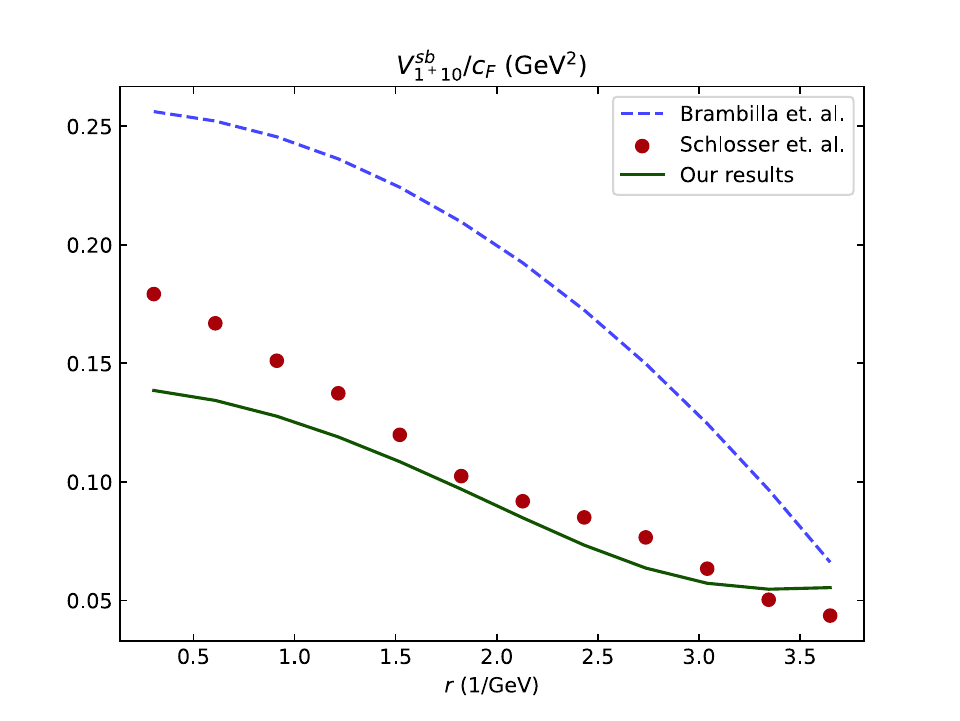} 
\caption{
Comparison of the $V_{1^+11}^{sa}(r)/c_F$ (left) and $V_{1^+10}^{sb}(r)/c_F$ (right) potentials in Ref. \cite{Soto:2020xpm} with our model (solid green), with lattice results from \cite{Schlosser:2025tca} (red dots), and with results from Table II of \cite{Brambilla:2019jfi} (dashed blue).
}\label{fig:V_relation}
\end{center}
\end{figure*}

After the previous version of this paper (v2) was published, lattice results on the spin-dependent potentials $V_{1^+\Lambda\Lambda'}^{sa}(r)$ and $V_{1^+\Lambda\Lambda'}^{sb}(r)$ appeared \cite{Schlosser:2025tca}. Hence,
we now include a comparison between our potentials, those from ref. \cite{Brambilla:2019jfi}, and the recent lattice computation of 
ref. \cite{Schlosser:2025tca} in Fig. \ref{fig:V_relation}. 
The potentials in \cite{Brambilla:2019jfi} relate to $V_{1^+11}^{sa}(r)$ and  $V_{1^+10}^{sb}(r)$ in \cite{Soto:2020xpm} by
\begin{gather}
\label{pot_relation}
    V_{1^+11}^{sa}(r)=V_{SK}(r)\\
   V_{1^+10}^{sb}(r) = V_{SK}(r) + V_{SKb}(r)r^2 \, .
\end{gather}
Recall that our $V^{sa}_{1^+11}$ has opposite sign to the one in Ref. \cite{Soto:2020xpm}.

We thank P. Heinrich, C. Schlosser, J. Tarr\'us, A. Vairo  and M. Wagner for discussions concerning this comparison, which led to identify several mistakes in the previous version (v2) of this paper.


\begin{appendices}
	\appendix
\section{Effective string theory calculation}
\label{string}
\begin{widetext}

The general expressions for the (heavy quark) spin dependent potentials (\ref{sdp32}), given in Ref. \cite{Soto:2020xpm}, reduce for $\kappa^p=1^+$ to two potentials \footref{sign}
,
\bea
\label{lattice}
\frac{V^{sa}_{1^+11}(r)}{m_Q}&=&\lim_{T\to \infty}\frac{g\,c_F}{m_Q T}\int_{-\frac{T}{2}}^{\frac{T}{2}} dt \frac{\langle B^\ast({\bf 0},\frac{T}{2}){ B^3}(\frac{\bf r}{2},t){ B}({\bf 0},-\frac{T}{2}) -B({\bf 0},\frac{T}{2}){ B^3}(\frac{\bf r}{2},t){ B^\ast}({\bf 0},-\frac{T}{2})
\rangle_\Box}
{\langle { B}({\bf 0},\frac{T}{2}){ B^\ast}({\bf 0},-\frac{T}{2})+{ B^\ast}({\bf 0},\frac{T}{2}){ B}({\bf 0},-\frac{T}{2})
\rangle_\Box }\\
\frac{V^{sb}_{1^+10}(r)}{m_Q}&=&\lim_{T\to \infty}\frac{g\,c_F}{4m_Q}\frac{ V_{\Pi_u}-V_{\Sigma_u^-}}{\sin \left( (V_{\Pi_u}-V_{\Sigma_u^-})\frac{T}{2}\right)}\int_{-\frac{T}{2}}^{\frac{T}{2}} dt
\frac{ \langle \left(B^\ast({\bf 0}, \frac{T}{2}){B}(\frac{\bf r}{2},t)-B({\bf 0}, \frac{T}{2}){B^\ast}(\frac{\bf r}{2},t)\right){ B^3}({\bf 0},-\frac{T}{2})\rangle_\Box }{\langle{ B}({\bf 0},\frac{T}{2}){ B^\ast}({\bf 0},-\frac{T}{2})+{ B^\ast}({\bf 0},\frac{T}{2}){ B}({\bf 0},-\frac{T}{2})
\rangle_\Box^{\frac{1}{2}} \langle {B^3}({\bf 0},\frac{T}{2}){ B^3}({\bf 0},-\frac{T}{2})
\rangle_\Box^{\frac{1}{2}}}\nn
\eea
\end{widetext}
where ${ B^i}({\bf r},t)$, $i=1,2,3$ are the three components of the chromomagnetic field, and
\bea 
B({\bf r},t)&=&B^1({\bf r},t)+iB^2({\bf r},t)\\
B^\ast({\bf r},t)&=&B^1({\bf r},t)-iB^2({\bf r},t)\, .\nn
\eea
${\bf {\hat r}}$ is taken in the $z$ direction.
The mapping of the operator insertions in the temporal Wilson lines onto EST operators was given in \cite{Perez-Nadal:2008wtr},
\bea
B(\pm{\bf r}/2,t)&\rightarrow&-i\sqrt{2}\Lambda'\partial_t\partial_z \varphi (\pm r/2, t)\nn\\
B^\ast(\pm{\bf r}/2,t)&\rightarrow&i\sqrt{2}\Lambda'\partial_t\partial_z \varphi^\ast (\pm r/2, t)\\
B^3(\pm{\bf r}/2,t)&\rightarrow&i\Lambda'''\partial_t\partial_z \varphi (\pm r/2, t)\partial_z \varphi^\ast (\pm r/2, t)+{\rm H.c.}\, ,\nn
\eea
and the mapping of the operator insertions in the middle of spacial Wilson lines at $t=\pm T/2$ onto string states in \cite{Oncala:2017hop},
\bea
{ B}({\bf 0},t) &\to & \pm i{\tilde \Lambda}^2\partial_0 \varphi ({ 0},t)\nn \\
 { B^\ast}({\bf 0},t) &\to & \mp i{\tilde \Lambda}^2\partial_0 \varphi^\ast ({ 0},t) \\
{ B^3}({\bf 0},t) &\to &  \pm {\tilde {\Lambda}}^{\prime 2}i
\partial_0 \varphi ({ 0},t) \partial_z \varphi^\ast ({ 0},t)+{\rm H.c.}\, ,
\nn
\eea
$t=\pm \frac{T}{2}$.  Note that neither $V^{sa}_{1^+11}(r)$ nor $V^{sb}_{1^+11}(r)$ depend on the values of ${\tilde \Lambda}^2$ or ${\tilde {\Lambda}}^{\prime 2}$. $V^{sa}_{1^+11}(r)$ is not sensitive to the sign ambiguity of the mapping either, however $V^{sb}_{1^+10}(r)$ is, which produces a sign ambiguity in it. Recall that $V_{\Pi_u}-V_{\Sigma_u^-}=-\frac{2\pi}{r}$ in EST. Once these substitutions are carried out in (\ref{lattice}), and the QCD vacuum averages $\langle\dots \rangle_\Box$ replaced by the string vacuum average $\langle\dots \rangle$,  a simple (tree level) calculation in EST leads to (\ref{Vhfl}) \cite{javier}.
\end{appendices}

{}

\end{document}